\documentclass[prl,aps,twocolumn,showpacs,amssymb]{revtex4}
\usepackage{color,graphicx,shortvrb}
\usepackage{amsfonts}
\usepackage{tabularx}
\usepackage{dcolumn}

\begin{document}
\title{Ultracold Molecule Production Via a Resonant Oscillating Magnetic Field}
\author{S.T.~Thompson, E.~Hodby, and C.E.~Wieman}
\address{JILA, National Institute of Standards and Technology and the University of Colorado, and the Department of Physics, University of Colorado, Boulder, Colorado 80309-0440}
\date{\today}

\begin{abstract}
A novel atom-molecule conversion technique has been investigated.
Ultracold $^{85}$Rb atoms sitting in a DC magnetic field near the 155~G Feshbach resonance are associated by applying a small sinusoidal oscillation to the magnetic field. There is resonant atom to molecule conversion when the modulation frequency closely matches
the molecular binding energy. We observe that the atom to molecule conversion efficiency depends
strongly on the frequency, amplitude, and duration of the applied
modulation and on the initial phase space density of the sample.
This technique offers high conversion efficiencies without the
necessity of crossing or closely approaching the Feshbach
resonance and allows precise spectroscopic measurements.
\end{abstract}
\pacs{03.75.Nt, 34.50.-s, 36.90.+f}
\maketitle

The efficient conversion of ultracold atoms to ultracold molecules
by time varying magnetic fields in the vicinity of a Feshbach
resonance is currently a topic of much experimental and
theoretical interest.  This particular conversion process lends
itself well to the formation of molecular Bose-Einstein
condensates (BECs) and atom-molecule superpositions\cite{Donley,Greiner,KetterleBEC}. These
Feshbach molecules and their creation process are also important
for understanding ultracold fermionic systems in the BCS-BEC
crossover regime because they are closely related to the pairing
mechanism in a fermionic superfluid that occurs near a Feshbach
resonance\cite{Holland,Timmermans,KetterleBCS,GreinerBCS}. Finally, Feshbach molecules
are interesting themselves because they are very weakly bound and
very large in spatial extent - comparable to the spacing between
atoms in the sample from which they were created\cite{Burnett}.

To date three Feshbach molecule creation techniques have been
demonstrated. Atom to molecule conversion was first directly
observed by applying very rapid (10's of $\mu$s) time dependent
magnetic fields to a $^{85}$Rb BEC in a Ramsey type
manner\cite{Donley}. The magnetic field was pulsed very close to
the Feshbach resonance which created a superposition of free atoms
and Feshbach molecules.  This technique was plagued by low
conversion efficiencies that were difficult to control.  It also lead to heating and loss of atoms from the atomic sample.

The most popular atom-molecule conversion scheme to date involves
slowly sweeping the magnetic field through a Feshbach resonance.
This has been demonstrated for both fermionic\cite{Regal,Hulet}
and bosonic\cite{GrimmCs,KetterleNafirst,Rempe,GrimmEuro,Hodby} atoms.
Although high conversion efficiencies have been observed in
degenerate fermi systems, high vibrational quenching rates near
the Feshbach resonance have lead to low conversion efficiencies
for BECs of bosonic atoms\cite{Rempe,KetterleNasecond}.  There are also problems caused by density dependent heating processes.  As the resonance is crossed, we
observe significant heating even in an uncondensed $^{85}$Rb sample
that is two orders of magnitude lower in density than typical BECs.  This is likely due to three body
recombination collisions. For adiabatic magnetic field sweeps, we have shown that the atom to molecule
conversion efficiency is solely determined by the phase space
density of the atomic sample\cite{Hodby}. Therefore, by limiting
the achievable phase space density, this heating is also limiting
the conversion efficiency.

A third atom-molecule conversion technique has been demonstrated
in two experiments with fermionic atoms\cite{Salomon,Grimm3}. This
technique utilizes the enhanced three body recombination collision
rates near a Feshbach resonance to efficiently associate atoms
into molecules.  Molecules were formed simply by holding a
degenerate fermi cloud of atoms for several seconds on the
positive scattering length side of a Feshbach resonance where a
weakly bound molecular state exists.  Conversion efficiencies as
high as 85\% have been reported\cite{Salomon}.  This technique
would not be useful in a bosonic system due to the comparatively
short lifetime of molecules formed from bosonic atoms.  The
longest observed lifetimes for molecules composed of bosonic atoms
are on the order of 10's of ms\cite{KetterleNasecond,Thompson}.

Low bosonic atom-molecule conversion efficiencies and the heating
observed when using both of these time dependent magnetic field
techniques has prompted us to investigate alternative conversion
methods. In this Letter we report on a novel atom-molecule
conversion method in which atoms are resonantly associated to form
molecules by applying a sinusoidally oscillating magnetic field
modulation. Note that this molecular formation process is different from typical photoassociation processes where a colliding pair of ground state atoms absorbs a photon to form an excited state molecule\cite{oldPA,Pillet,Stwalley}.  The photon \emph{adds} energy to the system, allowing a molecule to form.  In our case, photons from our oscillating magnetic field have the opposite effect - they cause the system to \emph{lose} energy by stimulating an atom pair to emit a very low frequency photon and thereby decay to a lower energy bound molecular state.  These experiments were motivated in part by work done by Regal
$et~al.$ where radio frequency photons were used to
\emph{dissociate} molecules formed from a degenerate two component
Fermi gas\cite{Regal}.

This technique circumvents problems of heating and
enhanced collision rates because it does not require crossing or
closely approaching the Feshbach resonance.  The maximum achievable
conversion efficiency for a given initial phase space density is
the same as was observed in the slow field sweep experiments\cite{Hodby}.
This new scheme also allows spectroscopy that is comparable or
superior to the Ramsey technique discussed in Ref.~\cite{Donley}.

A detailed description of the $^{85}$Rb experimental apparatus can
be found in Ref.~\cite{Wiemansetup}.  We use evaporative cooling
in a purely magnetic trap at a bias field of 162~G to produce
either degenerate or non-degenerate atomic samples.  Bose-Einstein
condensates used for these experiments contain 5000-10000 atoms
and a 50\% condensate fraction, and the non-degenerate clouds used
typically have 40000-200000 atoms at temperatures ranging from
20-80~nK.  The majority of these experiments were done with uncondensed samples. After producing an ultracold atomic sample, we ramp the
magnetic field from 162~G to a selected value between 156~G and
157~G in 5~ms (Feshbach resonance is located at 155.0~G).  We then
use the trapping coils to apply a sinusoidal modulation to the
magnetic field for 0-50~ms whose peak to peak amplitude ranges
from 130-280~mG and whose frequency is close to the molecular
binding energy as measured in Ref.~\cite{Neil}. We then slowly
ramp the field (in 5~ms) back to 162~G where the molecular
lifetime is 700~$\mu$s\cite{Thompson}. The field is held here for
about 10~ms to ensure that any molecules we've made decay,
ejecting their constituent atoms from the trap and our field of
view in the process\cite{Thompson}. The trap is then rapidly
turned off and we measure the number of atoms remaining using
absorption imaging. The loss of free atoms is negligibly small, so
any loss of atoms in this process must be due to molecule
production. We observe significant molecular formation that is
very dependent on modulation frequency, time, and amplitude and on
the initial phase space density of the sample.

As further confirmation that the observed loss is due to molecular formation and decay, we have done
a slightly modified version of this experiment.  After applying
the sinusoidal modulation to create molecules, the trap is
immediately turned off instead of ramping the magnetic field back
to 162~G where all of the molecules would quickly dissociate.  In
turning the trap off, we are sweeping the magnetic field through
the Feshbach resonance, converting any remaining molecules back
into atoms.  In this case, most of the original atoms are still
present in the absorption image and the slight loss we observe is
consistent with the molecular lifetime at the field at which the
molecules were created.

Figure 1 shows the observed atom loss as a function of
modulation frequency for three different modulation durations
(coupling times).  There is a clear resonant frequency. As the
coupling time increases, more atoms are converted into molecules.
Over the range of modulation amplitudes and coupling times we have
investigated, the width of the loss feature increases linearly
with peak loss to within our measurement uncertainty. For 37(2)\%
conversion at a field of 156.45~G the width is 1.88(16)~kHz while
for 17(2)\% conversion it is 84(12)~kHz.  By fitting the linewidth
versus conversion data to a straight line we find the width in the
zero percent conversion limit to be 0.2(2)~kHz.  We would expect
the width due to the spontaneous lifetime of the molecules to be
1/2$\pi$$\tau$ which, with $\tau$ = 11~ms, is 0.01~kHz.

\begin{figure}
\includegraphics[bb=85 113 573 503,clip,scale=0.45]{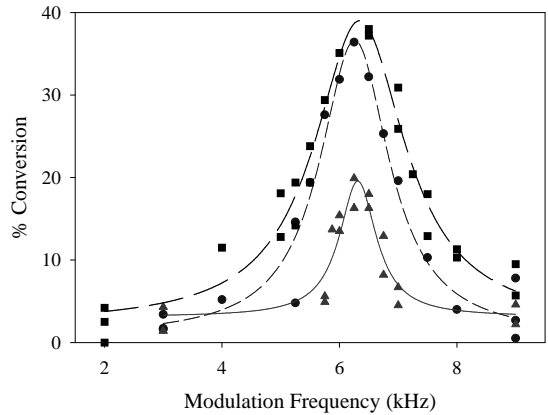}
\caption{Percentage of atoms converted to molecules at 156.45~G as
a function of modulation frequency for three different coupling
times.  The modulation amplitude was 130~mG.  The triangles are
the data and the solid line a Lorentzian fit for a coupling time
of 6~ms.  The circles and short dashed line are for a 25~ms
coupling time and the squares and long dashed line are for a 38~ms
coupling time.  The widths are 0.82(14)~kHz, 1.44(18)~kHz, and
1.88(16)~kHz respectively.  The peak positions for all three
curves agree within the 0.04~kHz fitting uncertainty.}
\end{figure}

We have investigated the dependence of the resonant frequency on
the value of the DC magnetic field. One would expect this to provide information about
the field dependent molecular binding energy.  The results of all of our
measurements are shown in Fig. 2.  The solid line is
the molecular binding energy that resulted from fitting our Ramsey
measurements described in Ref.~\cite{Neil} with an exact coupled
channels scattering calculation and the dashed lines represent the
uncertainty in that fit.  The solid circles are measurements made with
uncondensed thermal clouds that have been corrected for the temperature shift discussed below. The open squares are measurements made
with much denser, partially condensed clouds.

One would expect that the exact transition frequency would depend on the relative energy of the free atoms, and we see this as a temperature dependent shift in the measured resonant frequency as shown
in Fig. 3.  In order to form molecules, hotter, more energetic atoms require higher frequency photons to carry away their excess energy.  We have measured the resonant frequency at
155.44~G for clouds with temperatures of 20~nK, 50~nK and 80~nK.
Over this range the shift in the resonant frequency is linear with
a slope of 0.0126(4)~kHz/nK which corresponds to 0.60(2) $k_{B}T$ where
$k_B$ is Boltzmann's constant and $T$ is the temperature of the sample.  The thermal cloud data points shown in Fig. 2 have
been corrected to show the zero temperature resonant frequency.  Most of the error bars for these measurements overlap with the error bars on the binding energy curve but it is interesting that they are all consistently shifted to slightly higher frequency.  This is consistent with the peak of the Feshbach resonance being about 40~mG lower than the value found in Ref.~\cite{Neil}.

\begin{figure}
\includegraphics[bb=91 131 521 451,clip,scale=0.5]{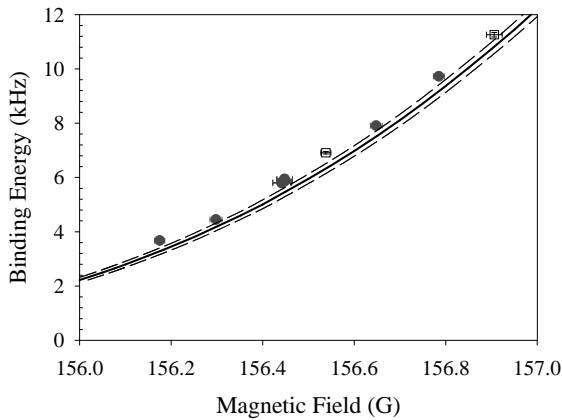}
\caption{Resonant frequency as a function of magnetic field.  The
solid line is from Ref.~\cite{Neil} and is the result of fitting our Ramsey measurements with a
coupled channels scattering calculation.  The  dashed line
indicates the uncertainty in that fit.  Measurements were made with both
condensed and uncondensed ($T$ = 20 - 80~nK) $^{85}$Rb clouds.  The open squares
indicate BEC measurements and the closed circles represent thermal
cloud measurements. The data has been corrected for the temperature dependent shifts discussed in the text.  The uncertainty in the measured frequency is represented by the width of the points.}
\end{figure}

\begin{figure}
\includegraphics[bb=202 158 578 442,clip,scale=0.5]{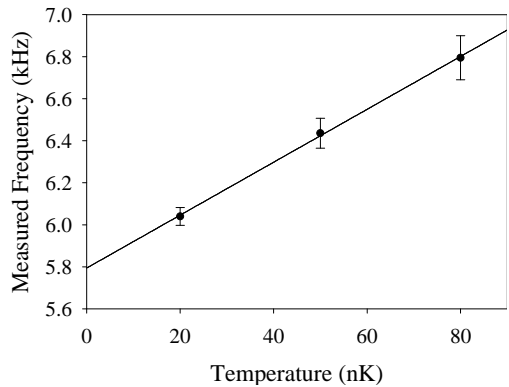}
\caption{Measured resonant frequency at 156.44~G as a function of
the temperature of the atomic sample.  Fitting the data to a
straight line gives a frequency shift equal to 0.60(2)~$k_{B}T$ where $k_{B}$
is Boltzmann's constant and $T$ is the temperature. }
\end{figure}

The two BEC measurements were carried out with clouds which had
50\% condensate fractions.  We found the results of BEC measurements to show larger statistical fluctuations which may be due to shot to shot fluctuations in the phase space density of the partially condensed samples.  The sample was not significantly
heated during the molecule production process and the condensate
fraction remained the same to within our level of uncertainty
(approximately 15\%).  Small corrections (0.03-0.06~kHz) due to the
5-10~nK thermal cloud fractions have been made to to the measurements shown
in Fig. 2.  We assumed that the thermal fraction and
condensate fraction have slightly different resonant frequencies
so that the frequency we measure is an average of the two.  The
measurement at 156.54~G lies above the best fit line for measurements made with lower density thermal clouds and hence may show evidence for a mean field shift to the binding energy.  A shift
consistent with this one was also observed using our Ramsey
technique as reported in Ref.~\cite{Neil}.

We have thoroughly investigated the resonant conversion efficiency
at 156.45~G as a function of coupling time.  Fig. 4a shows
that the conversion efficiency increases with coupling time until
a saturated value is reached.  Interestingly, this value is the
same as we would expect if we had instead ramped the magnetic
field adiabatically across the Feshbach resonance\cite{Hodby}.  In
both cases the maximum possible conversion is determined just by
the initial phase space density.  We have confirmed that this same
phase space limit on conversion efficiency applies at all magnetic
field values we have investigated.

Figure 4b shows the converted fraction in the short coupling time
regime.  We observe Rabi-like oscillations in the atomic
population at a frequency of about 2.5~kHz.  Note that the initial
peak to peak amplitude of these oscillations is only 6\% of the
total atom number, making this measurement rather difficult.  We
have mapped out the oscillations for slightly longer coupling
times and observe that, as one would expect, they damp out with a
time constant of 4~ms which is of the same order as the molecular
lifetime.  One question that remains is why the initial
oscillation amplitude (6\%) is much less than the maximum
converted fraction of 30\%.  In a simple two level system one would
expect a Rabi oscillation amplitude of 30\% because the maximum conversion would
occur during the first Rabi cycle.  Here, we observe a slow
build up of molecules over a time scale much slower than the Rabi
frequency.  It is unclear what determines the coupling strength
relevant to both these Rabi-like oscillations and the time scale
associated with reaching the saturated conversion regime.

\begin{figure}
\includegraphics[bb=157 68 481 557,clip,scale=0.6]{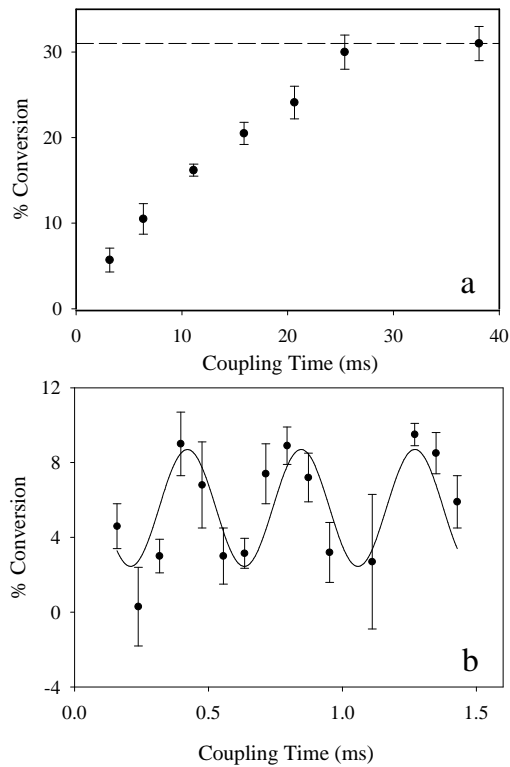}
\caption{(a) Percentage of atoms converted to molecules at
156.45~G (using 6.5~kHz resonant modulation frequency) as a function of coupling time.  The number of molecules
increases with coupling time until the conversion becomes
saturated.  (b) Oscillations in converted fraction as a function
of coupling time.  The oscillation frequency is approximately
2.5~kHz and the damping time is approximately 4~ms. }
\end{figure}

In summary, we have devised a new and very useful technique
for atom-molecule coupling in the vicinity of a Feshbach resonance
that offers superb control of the conversion process.  It allows
conversion efficiencies that are equal to those achieved by
ramping the magnetic field through the Feshbach resonance, but
avoids problems associated with working close to resonance such as
heating and rapid collisional loss.  We have observed Rabi-like oscillations between atomic
and molecular populations that damp out on the timescale of the
molecular lifetime.  This new technique also allows precise
spectroscopic measurements that are comparable to the results of
our Ramsey measurements in Ref.~\cite{Neil}.

In the future we hope to use this
new technique to produce and further study molecular BECs and
atom-molecule superposition states. We also plan to use it as a precise spectroscopic tool for investigating the atom-molecule
mean field shift close to the Feshbach resonance for high density
clouds.  It should also be possible to do a Ramsey type separated oscillatory field spectroscopy.  Oscillations between the atomic and molecular populations should occur as a function of the time between the two modulation periods.

We thank C.~Regal, D.~Jin, E.~Cornell and S.~Papp for helpful
discussions.  This work has been supported by ONR and NSF.

\end{document}